\documentstyle[epsf]{aa520}
\newcommand{\msun}{M$_{\odot}$}
\newcommand{\minit}{$M_{\rm init}$}

\newcommand{\hi}{H\,{\sc i}}

\newcommand{\ea}{et~al.~}

\newcommand{\psqc}{cm$^{-2}$}

\begin{document}
 
\title{The contribution of halo red giant mass loss\\ 
to the high-velocity gas falling onto the Milky Way disk}

\titlerunning{Halo red giant mass loss and HVC infall rate}

\author{Klaas S. de~Boer\inst{}
}

\institute{Sternwarte, Universit\"at Bonn, Auf dem H\"ugel 71, D-53121 Bonn, Germany
}

\date{Received 24 December 2003/ Accepted 10 February 2004}

\offprints{deboer@astro.uni-bonn.de}

\abstract{The origin of gas falling from the halo 
toward the disk of the Milky Way is still largely unclear. 
Here the amount of gas shed by the (older) halo red giants is estimated. 
The distribution of red giants (RGs) in the halo is not known 
but that of a subset of stars in the post RG phase, 
the sdB stars of the horizontal-branch (HB), is. 
Using the mid-plane density and $z$-distribution of sdB stars, 
the ratio of sdB stars to all HB stars, and the RG mass loss, 
the infall due to total mass lost by all halo RG stars at $z>1$ kpc 
is calculated. 
For the extended halo component 
$\dot{M}_{\rm halo\ RGs} \simeq 
  1.4 \cdot 10^{-5}$~\msun\ kpc$^{-2}$~yr$^{-1}$ 
while the thick disk component RGs contribute 
$\dot{M}_{\rm thick\ disk\ RGs} \simeq 
  5.4 \cdot 10^{-5}$~\msun\ kpc$^{-2}$~yr$^{-1}$, 
each with an uncertainty of a factor 4. 
The total rate of infall due to RG mass-loss is 
$\dot{M}_{{\rm RGs\ at\ }z>1\ {\rm kpc}} \simeq 
  7 \cdot 10^{-5}$~\msun\ kpc$^{-2}$~yr$^{-1}$, 
a sizeable fraction 
of the equally uncertain observed rate of infall of material. 
Since most of the RG stars in the extended halo are old, 
their mass loss is predominantly metal-poor, 
while that of the disk RGs is more metal-rich. 
The galactic fountain flow provides additional metal-rich infall 
and small galaxies being accreted contribute to the infall of gas as well. 

}

\maketitle

\section{The infall problem}

The detection at high galactic latitude of complexes of neutral hydrogen gas 
having their velocity directed toward the Solar vicinity (Muller \ea 1963)
showed that neutral gas clouds are falling toward the Milky Way disk. 
It was speculated that the gas might come from intergalactic space 
(Oort 1966), in which case it should be almost free of metals. 
The discovery of this influx of possibly pristine matter was embraced 
by modellers of the evolution of the Milky Way. 
The ``closed box'' evolution models predicted many more metal-poor G-dwarfs 
than actually are known to exist today 
(see reviews by Audouze \& Tinsley 1976 or Pagel 1997). 
Since that time the amount of infall needed in the evolution models 
was taken either as a free parameter or set equal to the amount of infall 
as can be estimated from the observed halo high-velocity clouds (HVCs). 

That rate of infall observed was rather uncertain 
because in the 1970s only part of the sky had been surveyed for \hi. 
In addition, 
the distance to the HVCs as well as intermediate velocity clouds (IVCs) 
was unknown (and is still largely unknown today; 
see review by Wakker \& van Woerden 1997). 
Knowledge of such distances is, of course, crucial for the calculation 
of the mass in HVCs and IVCs and thus of an infall rate. 
 
The infalling material\footnote{Essentially 
all gas seen at high galactic latitudes can be regarded as 
falling toward the Milky Way disk; 
to see that one has to correct each observed velocity for the 
velocity of the Sun in its orbit around the Milky Way. 
Moreover, any gas at large $z$-distance (thus being halo gas) 
having zero vertical velocity 
would blend in velocity with gas of the disk near the Sun. 
For more on the problem see, e.g., Kaelble \ea (1985).} 
turned out to contain metals, 
a fact clearly demonstrated by observations with the 
{\sc iue}, {\sc hst}, {\sc orfeus}, and {\sc fuse} spectrographs 
(see, e.g., Savage \& de~Boer 1981; Savage \& Sembach 1996; 
Richter \ea 2001; Collins \ea 2003). 
Yet the metal content varies from close to Solar to 1/10 or less 
(compare Richter \ea 1999 with Wakker \ea 1999). 
For a full list of IVCs and HVCs detected 
in metal absorption lines see Wakker (2001). 

These findings fueled the debate about the origin of the infalling gas. 
The models developed can be grouped into three categories. 
One category has the clouds as the cooling part 
of a galactic fountain-like flow (Shapiro \& Field 1976, Bregman 1980) 
and models for the kinematics of this flow show broad consistency 
with observed velocities (Bregman 1980, Kaelble \ea 1985, Wakker 1991). 
Another group sees some of the HVCs as material falling in from 
intergalactic space (Oort 1966, 1970), 
possibly even gas from within the Local Group (Blitz \ea 1999). 
In addition there is the gas of the Megallanic Stream, 
clearly strewn along the path of the Magellanic Clouds 
through the Milky Way halo, which falls toward the disk. 
Also gas from other caught satellite dwarf galaxies 
(c.f., Helmi \& White 1999) may be present. 

Also molecular hydrogen was detected in HVCs and IVCs 
(see Richter \ea 1999; Gringel \ea 2000; Bluhm \ea 2001; Richter \ea 2001), 
which would require the presence of dust. 
Recently, Evans \ea (2003) discovered dust in the globular cluster NGC~7078. 
This material will stay in the cluster until 
it can be stripped when the globular cluster on its orbit 
zippes through the Milky Way disk. 
Halo RG stars, however, would shed their gas and dust right into the halo. 

The amount of gas 
shed by red-giant (RG) stars in the halo of the Milky Way (MW) 
is proposed to be an essential component contributing to the infall. 
Its production rate is assessed, 
to be compared to the observational facts about infall available today.

\section{Age, mass, and metallicity of halo stars}

The MW halo is composed of stars belonging to the MW ab initio, 
most likely supplemented by stars accreted from satellite objects 
(e.g., the Sagittarius dwarf galaxy). 

Ab initio halo stars are as old as the globular clusters ($\simeq 13$ Gyr). 
They likely have a chemical composition 
similar to that of globular cluster stars. 
The main-sequence mass \minit\ of those stars being now RG was, 
assuming they are metal-poor ($Z=0.001$), between 0.92 and 0.82 \msun\ 
(see the evolution tracks by Schaller et al.\,1992). 

RG stars belonging to the MW thick disk 
will in part be younger and more metal-rich than the ab inito halo stars. 
Their \minit\ is larger, up to 1.5 \msun\ 
(or just up to 1.1 \msun\ for solar metallicity, $Z=0.02$; 
Schaller et al.\,1992). 

It has been speculated that some young stars are present in the halo. 
In high latitude blue star surveys stars were found that seemed to have 
(from Balmer line fits) main-sequence-like gravities. 
If these were indeed main-sequence stars, 
then one would have star formation in the halo. 
However, most such claims did not stand up to scrutiny. 
Their number is small anyway and their effect on 
overall halo star mass loss is negligible. 

Accretion of stars from satellite galaxies took place in the early phases 
of the MW (see the review by Freeman \& Bland-Hawthorne 2002). 
Those accreted satellites were rather metal-poor. 
Due to their disruption no new stars have formed in the debris since accretion 
so that their RGs have \minit\ in the same range 
as those proper of the MW halo. 
However, more recent accretion may have contributed more massive stars, 
but their fraction of the total halo mass is likely very small and so 
will be ignored. 

\section{Calculating the mass lost by halo Red Giants}

To calculate the total mass lost by RG stars, 
the mass loss, the spatial distribution of RGs in the halo, 
and the total number of RGs must be known. 
The RG star distribution is, however, essentially unknown and therefore must 
be derived in an indirect manner. 
The sdB stars serve the purpose. 
They are a subclass of HB stars and HB stars have RG stars as progenitors. 
The various parameters needed for the study are described in the 
following subsections. 
The RG mass loss will be discussed in Sect.\,\ref{masslost}, 
the spatial distribution of sdB stars is described in 
Sect.\,\ref{numberofstars} 
while the ratio of sdB stars to all HB stars is discussed 
in Sect.\,\ref{sdbhbratio}.
Finally, the mean RG mass loss rate is calculated in Sect.\,\ref{masslossrate} 
and the total mass lost is calculated in Sect.\,\ref{totalloss}.

\subsection{Mass lost by one RG star}
\label{masslost}

Red giants evolve and lose mass and, if $M_{\rm init} <  1.5$ \msun, 
end up as horizontal-branch (HB) stars. 
HB stars are well defined objects in the core He burning phase 
having a He core of $\simeq0.5$ \msun\ 
surrounded by a hydrogen shell ranging from  
at most 0.02 \msun\ (the very blue sdB and the BHB stars) 
via the HBA and RR Lyr stars 
to those with a shell of up to 0.4 \msun\ (the red HB (RHB) stars). 
The end product of the evolution of RGs is thus an HB star with 
$M_{\rm end} \simeq0.5$ \msun\ (BHB star) to 
$M_{\rm end} \simeq0.9$ \msun\ (RHB star). 
For the (old) halo RGs $M_{\rm init} < 1.0$ \msun. 

Combining the values given above for \minit\ and $M_{\rm end}$ 
(the lowest \minit\ and the highest $M_{\rm HB}$: 
0.82 and 0.9 \msun, respectively; 
the highest \minit\ and the lowest $M_{\rm HB}$: 1.0 and 0.55 \msun) 
shows that the total range possible for the amount lost by halo RGs is 
$M_{\rm lost}  =$ 0.0 to 0.45 \msun .
A reasonable {\sl mean} for the mass lost by metal-poor stars is 
\begin{equation}
\overline{M_{\rm lost}} = M_{\rm init} - M_{\rm end} \simeq 0.3 \ \ 
[{\rm M}_{\odot}] \ \ {\rm per \ star}. 
\label{eqmass1}
\end{equation} 

\subsection{Number and distribution of halo HB stars}
\label{numberofstars}

Several studies exist about the distribution of stars in the halo. 
Following earlier work, Chiba \& Beers (2000) derived 
equidensity contours for the halo distribution 
from the kinematics of solar neighbourhood metal-poor ([Fe/H]$<$$-1$) stars. 
One important result from their work is that 
the density distribution in the halo of the subsample with 
$-1.6$$<$[Fe/H]$<$$-1.0$ 
is not very different from that of the poorest subsample ([Fe/H]$<$$-1.8$), 
albeit somewhat more flattened. 
Chiba \& Beers give diagrams with the spatial distribution of stars 
in $R$ and $z$. 
Their distribution can, in the solar neighbourhood, be represented 
by an exponential with $h_z \simeq 20$~kpc.

We need, however, to know the distribution of just the RG stars. 
This is observationally difficult 
mostly because RGs cover such a large range in $M_V$. 
Stars in the next evolutionary stage, the HB stars, 
are representative for the RG distribution. 
Of those, the distribution of the sdB stars is well studied. 
We need to know the midplane density and the vertical distribution. 

Several studies exist aiming at finding the spatial distribution 
of sdB stars perpendicular to the MW dsik. 
In one study, the disk region was searched for these stars 
to arrive at a space density of 
$n(0)_{{\rm disk\ sdB}} \simeq 2 \cdot 10^{-6}$ pc$^{-3}$ (Downes 1986). 
Other studies explored well defined high-latitude fields, 
attempting to obtain a ``complete sample'' of sdB stars out to some distance 
and then fit a vertical distribution function (mostly an exponential) 
to arrive at a mid-plane density and a scale height. 
Examples of such studies are Heber (1986) and Moehler \ea (1990).
These led to scale heights of $h_{\rm sdB} \simeq 200$~pc. 
Villeneuve \ea (1995) used an all sky sample 
and a ``$V/V_m$ test'' (Schmidt 1968), 
arriving at mid-plane densities and scale heights as well. 
A thorough appraisal of such work and earlier studies 
was made by Villeneuve \ea (1995), concluding that 
$n(0)_{{\rm disk\ sdB}} = 3 (\pm 1) \cdot 10^{-7}$ pc$^{-3}$ 
and $h_{\rm sdB} = 600 (\pm 200)$~pc. 
Their midplane density will be used below. 

Not affected by problems of completeness of those studies is the 
derivation of the scale height from the $z$-distance statistics of a large 
sample of galactic orbits (calculated using a galactic potental). 
In this manner, Altmann \ea (2004) arrived at the existence of two 
populations of sdB stars, a halo one with $h_{\rm halo\ sdB}\simeq 7$~kpc 
and a disk one with $h_{\rm disk\ sdB}\simeq 0.9$~kpc. 
No absolute midplane density can be derived from their data, 
but they estimate that the midplane density ratio of halo 
and thick disk sdBs is 
$n(0)_{{\rm halo\ sdB}} = 0.0125 (\pm 25\%)\times n(0)_{{\rm disk\ sdB}}$. 

Note that on the direct observational side, 
an irregular halo HB star distribution out to 45 kpc 
is seen in the SDSS data (Yanny \ea 2000).

 From these parameters one can calculate the number ratio of sdB stars in 
each of these populations. 
Integration over the exponential distributions 
leads to  
\begin{equation}
 N_{\rm halo\ sdBs}/N_{\rm disk\ sdBs} = 
  \frac{n(0)_{{\rm halo\ sdB}} \cdot h_{({\rm halo\ sdB})}}
       {n(0)_{{\rm disk\ sdB}} \cdot h_{({\rm disk\ sdB})}}  
\end{equation}
so that, with the numbers of Altmann \ea (2004), 
the total number ratio of halo to disk sdB stars is 0.07. 

Combining the Villeneuve \ea (1995) mid-plane density 
with the halo to disk ratio from Altmann \ea leads to 
$n(0)_{\rm halo\ sdB} = 3.7 \cdot 10^{-9}$~pc$^{-3}$ $= 3.7$~kpc$^{-3}$. 

The total number of sdB stars in the halo follows from integration over 
the $z$-distribution 
$N= \int_{z_{\rm b}}^{z_{\rm t}} n(0)e^{-z/h_z} {\rm d}z$,
with $z_{\rm b}$ and $z_{\rm t}$ the bottom and the top of 
the layer of integration. 
Thus  
\begin{equation}
N_{\rm sdB} = n(0)_{\rm sdB}  h_z \ (e^{-z_{\rm b}/h_z}-e^{-z_{\rm t}/h_z}) 
      = n(0)_{\rm sdB}  h_z \ g
\label{eqnsdbz}
\end{equation}
in which $g$ is the integral's value. 

For the current analysis 
the rather well-defined Villeneuve \ea (1995) mid-plane density, 
the ratio of the space density of halo to disk sdB stars, 
and the respective scale heights 
will be adopted to represent the extended halo RG distribution. 
Both the halo and thick disk components are considered. 
For the thick disk $h_z = 0.9$~kpc from Altmann \ea (2004) and 
$n(0)_{\rm disk\ sdB} =3\cdot10^2$~kpc$^{-3}$ from Villeneuve \ea (1995). 
For the halo the values $h_z = 7$~kpc from Altmann \ea (2004)
and $n(0)_{\rm halo\ sdB} =3.7$~kpc$^{-3}$ (see above) will be used. 
The integration is from $z=1$~kpc to infinity and 
$g$ thus gives the fraction of stars of each population above $z=1$~kpc. 
For the halo sdBs $g= 0.87$ and for the (thick) disk sdBs $g= 0.33$.

\subsection{The ratio of sdB stars to all HB stars}
\label{sdbhbratio}

A further important parameter to estimate is the number ratio of 
sdB stars to all HB stars  
$  f_{\rm all\ HB/sdB} = n_{\rm all\ HB}/n_{\rm sdB}$.

One can look at CMDs of globular clusters and count the blue HB stars 
(within some colour range) in relation to the total number of HB stars. 
Here the problem is to define which globular cluster would be typical 
for the halo population. Such a cluster does not exist. 
Alternatively, one could 
add up the CMDs of several globular clusters each representing 
some part of the history and characteristics of the halo population. 
Such a procedure leads to a ratio of blue HB stars to all HB stars of 
crudely a factor of 10. 
However, 
the sdB stars are in CMDs rather on the ``vertical part'' of the HB, 
and this colour and brightness range is observationally still poorly sampled 
(but see, e.g., Rosenberg \ea 2000).  

A different approach is to relate the so-called ``birth rate'' of sdB stars 
in the MW disk to that of, e.g., the White Dwarfs. 
The birth rate is in a steady state the ratio of the observed space density 
divided by the phase life of such stars. 
For the sdB phase life one can use $t_{\rm HB}$, 
since the phase life of all HB stars covers the rather small range 
of 0.8 to 1.2 $10^8$~yr 
for all types (see Charbonnel et al.\,1996). 
Comparing the sdB star birth rate, $n_{\rm sdB}/t_{\rm HB}$, 
with that of the WDs, Villeneuve \ea (1995) 
find that the sdB stars make up just about 1\% of all stars becoming WD. 
Since WDs are the end product only of those stars that go through the HB phase, 
the number looked for, the number ratio of sdB stars to all HB stars,  is 
\begin{equation}
  f_{\rm all\ HB/sdB} = n_{\rm all\ HB}/n_{\rm sdB} = 100\ \ \ .
\label{eqfrac}
\end{equation}

The two approaches chosen differ in result by a factor of 10. 
The number through the birth rate is, 
given the method of derivation, much more likely to be accurate.

\subsection{Mass loss rate of one halo RG star}
\label{masslossrate} 

To calculate the total mass lost by the RG halo stars per unit of time 
one may attempt to define the length of the RG mass loss phase. 
Noticeable RG mass loss takes place only in the last 1\% of the 
evolution time from the main sequence up to the He-flash (end of the RG phase) 
according to the models of 
Schaller et al.\,(1992) and Charbonnel et al.\,(1996). 
This slow mass loss amounts to about 10\% \minit. 
However, the bulk of the mass loss occurs at the tip of the red giant branch 
and is episodic (Origlia \ea 2002), 
mostly in relation to the He flash. 
Defining an average all halo RG mass loss rate with little knowledge 
about the RG mass loss process is nigh to impossible. 

A much simpler approach is to assume that also 
the population of the stellar halo is in a steady state, 
i.e., the number of HB stars currently existing 
is representative of the number of HB stars over a considerable time. 
The phase life of HB stars, $t_{\rm HB}$ (the phase of core-He burning), 
is well known. 
As mentioned above, it is 0.8 to 1.2 $10^8$~y (Charbonnel et al.\,1996). 
The shorter time in this range is for metal poor stars, 
the longer for solar metallicity stars.
Thus each HB star (all of them have been RG star) 
represents an RG star with its mass shed. 

The {\sl spatially and time averaged mass loss rate} 
of one RG star thus must be 
\begin{equation}
 \overline{\dot{M}_{\rm one \ RG}} = 
\overline{M_{\rm lost}} \ / \ t_{\rm HB} \ \simeq 0.3/10^8 
   \ \ \  {\rm M}_{\odot} \ {\rm yr}^{-1} ,
\label{eqlossrate}
\end{equation}
with $\overline{M_{\rm lost}}$ from Eq.\,\ref{eqmass1} 
and $t_{\rm HB}$ from above. 

\subsection{The total halo RG mass loss and infall rate}
\label{totalloss} 

The total RG mass-loss rate projected onto a unit area of the Milky Way disk 
from one hemisphere follows combining 
Eqs.\,\ref{eqnsdbz}, \ref{eqfrac}, and \ref{eqlossrate} 
given above into
$ \dot{M}_{\rm halo\,RGs} =  
   N_{\rm sdB} \times f_{\rm all\,HB/sdB} \times 
   \overline{\dot{M}_{\rm one\,RG}} 
   \ \ {\rm M}_{\odot}\  {\rm kpc}^{-2}\ {\rm yr}^{-1} . $
This mass loss is calculated for the halo 
and thick-disk components separately. 
To account for both hemispheres, its value has to be multiplied 
by a factor of 2 
so that the total RG mass-loss rate projected onto MW disk
\begin{equation}
\dot{M}_{\rm RGs} = 
 2\times  n_{\rm 0\ sdB} \times h_z \times g \times f_{\rm all\ HB/ sdB} \times 
  \overline{{M}_{\rm lost}} / t_{\rm HB}
\label{overallloss}
\end{equation}
in units of \msun\ kpc$^{-2}$ yr$^{-1}$.

\subsubsection{Extended Halo RG mass loss}

For the extended halo component the values determined above, 
using $f_{\rm all\,HB/sdB}=100$, 
lead with Eq.\,\ref{overallloss} to 
$\dot{M}_{\rm halo\ RGs} = 1.37 \cdot 10^{-5}$~\msun\ kpc$^{-2}$~yr$^{-1}$.

In addition, a globular cluster loses on average 
about $10^{43}$ atoms s$^{-1}$ (de Boer 1985) 
equalling a total of $4 \cdot 10^{-7}$ \msun\ kpc$^{-2}$~yr$^{-1}$ 
for the $\simeq 150$ Milky Way clusters. 
This GC RG mass loss adds only 0.5\% to the field star value. 
The total is thus 
\begin{equation} 
\dot{M}_{\rm total\ halo\ RGs} \simeq 
  1.4 \cdot 10^{-5} \ {\rm M}_{\odot}\ {\rm kpc}^{-2}\ {\rm yr}^{-1} \ \ \ .
\label{haloloss}
\end{equation}
This mass loss is metal-poor. 

\subsubsection{Thick disk contribution to halo RG mass loss}

In addition, the RG stars of the thick disk lose mass. 
Performing the same calculation as for the extended halo component 
one arrives with Eq.\,\ref{overallloss} for the thick disk at 
\begin{equation} 
\dot{M}_{\rm total\ thick\ disk\ RGs} \simeq 
  5.4 \cdot 10^{-5} \ {\rm M}_{\odot}\ {\rm kpc}^{-2}\ {\rm yr}^{-1} \ \ \ ,
\label{diskloss}
\end{equation}
mass clearly lost in the lower halo, at $1 < z < 3$ kpc. 
It has a metallicity between somewhat below solar 
and low metallicity like that of the halo. 

\subsubsection{Grand total of halo RG mass loss, uncertainties}

The sum of both mass loss rates, 
being the mass lost by all RG stars at $z>1$ kpc, 
material which falls toward the MW disk, equals 
$\dot{M}_{\rm total\ RG} \simeq 7\cdot 10^{-5}$ \msun\ kpc$^{-2}$ yr$^{-1}$. 

The uncertainties in the mass loss rates $\dot{M}_{\rm RGs}$ 
(Eqs.\,\ref{haloloss} and \ref{diskloss}) can be estimated. 
The value of  $n_{\rm 0\ sdB\,disk}$ has an uncertainty of 30\%, that of 
the ratio of disk to halo sdB stars of about 25\%. 
The uncertainty in $h_z$ is 15\% for the halo and 10\% for the disk 
population. 
That of $g$ (depending on $h_z$) is 3\% for the halo population 
and perhaps up to 15\% for the disk one. 
The ratio $f_{\rm all\ HB/ sdB}$ from the birth rates is 100, 
from counts of stars in CMDs it is 10. 
The first procedure is more accurate but still may be unceratin 
by a factor of 2.
The mass lost per star ($M_{\rm lost}$) may be different 
by at most 50\% from the adopted value.
The mass loss rate follows from halo steady state 
and the well-defined HB phase life. 
In all, this means a total uncertainty of a factor $\simeq 5$ 
(for each population). 

\section{Observed infalling gas}

The infall rate of halo gas through HVCs has been calculated 
by a few authors. 
Oort (1966, 1970) derived a value of 
$\simeq 3 \cdot 10^{18}$ atoms\ cm$^{-2}$ Myr$^{-1}$ from early 21 cm data. 
This translates to close to $10^{-2}$ \msun\ kpc$^{-2}$ yr$^{-1}$.

Wakker \ea (1999) calculate the infall rate for just Complex C 
(if at $z=5$ kpc) as $\simeq 3 \cdot 10^{-3}$ \msun\ kpc$^{-2}$ yr$^{-1}$. 
Complex C is the biggest HVC present today covering $\simeq4$\% of the sky. 
Considering the entire sky one can define a filling factor, $f_{\rm fill}$, 
for high-velocity gas. 
Its value depends on the column density limit chosen. 
A reasonable value is 
$f_{\rm fill} \simeq0.1$ for $N_{\rm H}> 10^{18.5}$ \psqc\ 
(see the reappraisal of the all-sky coverage by Wakker 2004).
Thus a very uncertain extrapolation of Complex C to the entire sky 
(surface area covered, area of MW, filling factor) 
might mean for all HV gas 
$M_{\rm infall\ HVC} \simeq 1 \cdot 10^{-4}$ \msun\ kpc$^{-2}$ yr$^{-1}$. 

Such calculations are heavily biased by choices made for 
distances, space velocities, and densities of halo clouds. 
The velocities detected are essentially LSR velocities, 
and a distinction in HVCs and IVCs remains arbitrary with respect to 
the true space motion of the gas complexes, 
also because there are clear signs that halo gas does not co-rotate 
with the disk (de~Boer \& Savage 1983, Kaelble \ea 1985). 
The velocities themselves do not reveal much about the location of the gas. 
Moreover, the Magellanic Stream is not part of the general infall 
but can be accounted for as a special case with well-known origin. 

Clearly the ``observed'' infall rate is very uncertain because of the 
imponderable space velocities and distances. 
As long as no better data (especially distances) are available, 
there is little hope for improving the estimates 
of the amount of infall of gas from the halo. 

\section{Discussion}

The RG stars in the halo of the Milky Way lose a considerable amount of mass 
which is calculated in units equivalent to infall. 
The amount derived is a sizeable fraction of the amount estimated for 
the observed H\,{\sc i} gas falling in from the halo. 
Both estimates, of the total RG mass loss and of the detected infall, 
are quite uncertain. 

Yet, the metallicities as derived from the various UV absorption 
spectroscopy studies are of relevance for the discussion. 

The intermediate velocity gas is relatively metal-rich 
(Savage \& de Boer 1981, Richter \ea 2001) 
while the gas of Complex~C is metal-poor (Richter \ea 2001, Collins \ea 2003). 
The intermediate velocity gas is thought to be not far from the disk 
and could include gas of the galactic fountain. 
As shown above, also the RG stars of the lower halo will contribute 
gas with metallicity probably not far below solar. 

The more distant parts of the halo are the regions 
where metal-poor halo RG stars lose their mass. 
It is to be expected that the metal-poor gas shed at such distances will, 
after some time, condense and assemble in denser pockets 
which eventually fall toward the disk. 
Given the origin at larger $z$, a higher velocity of the downflow would 
be expected after an appropriate time. 

Other sources of material will contribute gas to the infall, too. 
The galactic fountain contributes metal-rich gas, 
while a portion of infalling gas is metal-rich Magellanic Stream gas 
and perhaps some gas from other accreted dwarf galaxies 
of unknown but likely lower metallicity. 
These three sources (RG stars, galactic fountain, dwarf galaxies) 
eliminate the need for substantial infall of ``pristine'' (zero metal) matter 
from intergalactic space.  

Note that in the past the halo RGs were stars with larger \minit. 
Then $\overline{M_{\rm lost}}$ must have been larger than today 
(see Weidemann 2000). 
But given the accepted shape of the initial mass function, 
their number must have been smaller and the total infall from RG mass loss 
was therefore not much different from that of today. 
Investigation of such effects is, given all uncertainties, 
beyond the scope of this paper. 

\acknowledgements
I thank the referee Bart Wakker for probing questions about this research 
which stimulated an extension of the original analysis, 
Uli Heber for enlightening discussions, 
and Martin Altmann, Michael Hilker, and Philipp Richter 
for critical readings of the manuscript.

\end{document}